\documentclass[preprints,article,accept,moreauthors,pdftex,10pt,a4paper]{Definitions/mdpi} 
\firstpage{1} 
\pubvolume{xx}
\issuenum{1}
\articlenumber{5}
\pubyear{2018}
\copyrightyear{2018}
\history{}

\usepackage{makecell}  %
\usepackage{siunitx}   %
\usepackage{amssymb}

\Title{Two-photon vibrational transitions in \OOp~as probes of variation of the proton-to-electron mass ratio}

\Author{Ryan Carollo \orcidA{}%
, Alexander Frenett \orcidB{} %
 and David Hanneke *\orcidC{}}

\AuthorNames{Ryan Carollo, Alexander Frenett and David Hanneke}

\address[1]{%
Physics \& Astronomy Department, Amherst College, Amherst, Massachusetts 01002, USA
}

\corres{Correspondence: dhanneke@amherst.edu; Tel.: +1-413-542-5525}

\abstract{Vibrational overtones in deeply bound molecules are sensitive probes for variation of the proton-to-electron mass ratio $\mu$. In nonpolar molecules, these overtones may be driven as two-photon transitions. Here, we present procedures for experiments with \OOp, including state-preparation through photoionization, a two-photon probe, and detection. We calculate transition dipole moments between all \Xstate~vibrational levels and those of the \Astate~excited electronic state. Using these dipole moments, we calculate two-photon transition rates and AC-Stark-shift systematics for the overtones. We estimate other systematic effects and statistical precision. Two-photon vibrational transitions in \OOp~provide multiple routes to improved searches for $\mu$ variation.}


\newcommand{\Be}{Be$^+$}
\newcommand{\OOp}{\ensuremath{\textrm{O}_2^+}} %
\newcommand{\Xstate}{\ensuremath{X\,^2\Pi_g}}
\newcommand{\XstateHalf}{\ensuremath{X\,^2\Pi_{g,\frac{1}{2}}}}
\newcommand{\Astate}{\ensuremath{A\,^2\Pi_u}}
\newcommand{\dstate}{\ensuremath{d\,^1\Pi_g}}
\newcommand{\SigmaState}{\ensuremath{1\,^2\Sigma_u^+}}
\newcommand{\maxA}{\ensuremath{\quad v_{A}^{(\textrm{max})}}} %

\newcommand{\ket}[1]{\ensuremath{|#1\rangle}}
\newcommand{\braOket}[3]{\ensuremath{\langle#1|#2|#3\rangle}}

\begin{document}
\section{Introduction}

Even simple molecules contain a rich set of internal degrees of freedom. When these internal states are controlled at the quantum level, they have many applications in fundamental physics~\cite{demilleScience2017,safronovaRMP2018} such as searches for new forces~\cite{salumbidesPRD2013}, investigation of parity~\cite{demillePRL2008a,quackARPC2008} and time-reversal~\cite{cairncrossPRL2017,acmeNature2018} symmetries, or searches for time-variation of fundamental constants~\cite{carrNJP2009,chinNJP2009,jansenJCP2014}. %
 Experiments with molecular ions~\cite{bresselPRL2012,germannNaturePhys2014,biesheuvelNComms2016,cairncrossPRL2017,calvinJPCL2018} are already at the forefront of these scientific questions, taking advantage of the long interrogation times allowed in trapped systems.

Because some degrees of freedom involve motion of the nuclei themselves, molecules possess the potential to probe for changes in the proton mass relative to the electron mass. This mass ratio, $\mu = m_p/m_e$, is predicted to change over time by several extensions to the standard model. Some models of quantum gravity include extra spatial dimensions or new scalar fields and suggest a drift in $\mu$ on cosmological timescales and continuing to the present day~\cite{uzanRMP2003,calmetMPLA2015}. Ultralight dark matter could cause $\mu$ to oscillate at a frequency set by the mass of the dark-matter particle~\cite{stadnikPRL2015a,arvanitakiPRD2015}; topological dark matter could cause transient changes in $\mu$~\cite{dereviankoNPhys2014}. Models typically predict that variation in $\mu$ should be approximately 40 times larger than corresponding changes in the fine structure constant $\alpha$~\cite{uzanRMP2003}.

Current limits on present-day variation in $\mu$ come from atomic clock experiments and find that $\dot{\mu}/\mu \lesssim 10^{-16}~{\rm yr}^{-1}$~\cite{godunPRL2014,huntemannPRL2014}. The sensitivity to $\mu$ in these experiments is through the hyperfine structure of cesium. Linking the hyperfine frequency to the nuclear mass requires a model of the cesium nuclear magnetic moment~\cite{flambaumPRC2006}. The reliance on cesium clocks also means that atomic techniques are nearing their feasible limits, as the cesium microwave clock has been surpassed in stability by optical atomic clocks~\cite{ludlowRMP2015}. These optical clocks are based on electronic -- not hyperfine -- transitions, so they have little sensitivity to $\mu$ variation.

The vibration and rotation of molecules provide a model-independent means to search for variation in $\mu$~\cite{schillerPRA2005,chinPRL2006,flambaumPRL2007,carrNJP2009,chinNJP2009,jansenJCP2014}. The current limit for a molecular experiment is ${\dot{\mu}/\mu = (-3.8\pm5.6)\times10^{-14}~{\rm yr}^{-1}}$, which is based on a rovibrational transition in SF$_6$ and was conducted in a molecular beam~\cite{shelkovnikovPRL2008}. Several proposals exist for next-generation searches in diatomic molecules~\cite{schillerPRA2005,chinPRL2006,flambaumPRL2007,demillePRL2008,zelevinskyPRL2008,kajitaJPB2011,kajitaPRA2014,hannekePRA2016,kajitaPRA2017,kokishArxiv2017,stollenwerkAtoms2018}. In this manuscript, we provide additional support for the use of the \OOp~molecule~\cite{hannekePRA2016,kajitaPRA2017} in such a search.

Searches for a change in $\mu$ usually involve monitoring the energy difference $hf$ between two energies with different $\mu$ dependence, $hf = E^\prime(\mu) - E^{\prime\prime}(\mu)$. The fractional change in $\mu$ is then related to an absolute frequency shift $\Delta f$ through
\begin{equation}
	\frac{\Delta\mu}{\mu} = \frac{1}{\mu}\left(\frac{\partial f}{\partial\mu}\right)^{-1}\Delta f = \left[\frac{\partial f}{\partial(\ln\mu)}\right]^{-1}\Delta f = \frac{\Delta f}{f_\mu}.
\end{equation}
In the last equation, we have defined the absolute sensitivity
\begin{equation}
	f_\mu \equiv \mu \frac{\partial f}{\partial\mu} = \frac{\partial f}{\partial (\ln\mu)} \label{eq:absSens}
\end{equation}
as the absolute frequency shift for a given fractional shift in $\mu$. This quantity is sometimes called the absolute enhancement factor.

Both vibrational and rotational degrees of freedom have sensitivity to $\mu$ variation because both involve the nuclei moving. In general, vibrational sensitivity scales as $\mu^{-1/2}$ and increases linearly with vibrational quantum number $v$. Rotational sensitivity scales as $\mu^{-1}$ and increases as $J(J+1)$. Anharmonicity and centrifugal distortion reduce the sensitivity for higher vibrational or rotational states~\cite{herzbergVolI1950,beloyPRA2011,pastekaPRA2015}. In general, vibrational transitions provide a favorable route to measure $\Delta \mu$. They typically have much higher absolute frequencies compared to rotational transitions. Rotational changes are also constrained by selection rules to small $\Delta J$, whereas it is possible to drive vibrational overtones with large $\Delta v$.

The \OOp~molecule is homonuclear and thus intrinsically nonpolar. Nuclear symmetry eliminates half the rotational states in any electronic manifold. The nuclei of \OOp~are spin-0, such that its \Xstate~ground state has only symmetric rotation states (see for example ref.~\cite[Sec.\,V2c]{herzbergVolI1950}). The absence of opposite-parity states forbids electric-dipole ($E1$) transitions between vibrational or rotational levels within the same electronic state. This absence suppresses many electric-field-related systematic effects without the need to average over multiple transitions~\cite{schillerPRL2014,kokishArxiv2017}. However, it means that any transition between vibrational states within \Xstate~requires some higher-order process.

One option would be to use the electric quadrupole ($E2$) coupling between states. The $v=0\leftrightarrow 1$ transition has been observed in the N$_2^+$ nonpolar molecular ion~\cite{germannNaturePhys2014}. Quadrupolar overtone transitions in \OOp~have been proposed~\cite{kajitaPRA2017}. However, direct driving of such transitions is hampered by the very small quadrupole moments for large $\Delta v$.

Alternatively, electric-dipole coupling to other excited electronic states provides a mechanism for driving two-photon transitions~\cite{hilicoJPB2001,zelevinskyPRL2008,karrJMS2014,kajitaPRA2014}. In this manuscript, we lay out potential experiments using such two-photon vibrational overtones. We calculate transition rates and systematic shifts for these experiments using existing spectroscopic data~\cite{coxonJMS1984,songJCP1999,songJCP2000}. Many of these overtones are accessible with conventional laser technology. We find the systematics for these experiments allow matching the $\dot{\mu}/{\mu} < 6\times10^{-14}~{\rm yr}^{-1}$ molecular limit~\cite{shelkovnikovPRL2008} in a trapped ensemble of multiple molecular ions. With fewer ions and a quantum-logic scheme the system is capable of measurements several orders of magnitude below the present best limit. %
 
\section{Experimental procedures}

Despite the appealing properties of the two-photon transitions between vibrational states, there are experimental challenges with state preparation and detection. Because \OOp is nonpolar and has non-diagonal Franck-Condon factors, it is not amenable to state preparation by optical pumping~\cite{staanumNPhys2010,schneiderNPhys2010,lienNatureComm2014}. For large numbers of molecules, as might be appropriate in preliminary investigations, one can create state-selected ions through photoionization. Detection of a transition can proceed by state-selective dissociation followed by analysis of the mass of the remaining ions. For small numbers of molecules, as might be appropriate in high-precision cases, one could pump and project molecules into an initial state by use of an auxiliary atomic ion and quantum control techniques~\cite{chouNature2017}. Non-destructive detection could also rely on quantum logic~\cite{schmidtScience2005,mur-petitPRA2012,wolfNature2016,chouNature2017}.

\begin{figure}%
\includegraphics[width=\columnwidth]{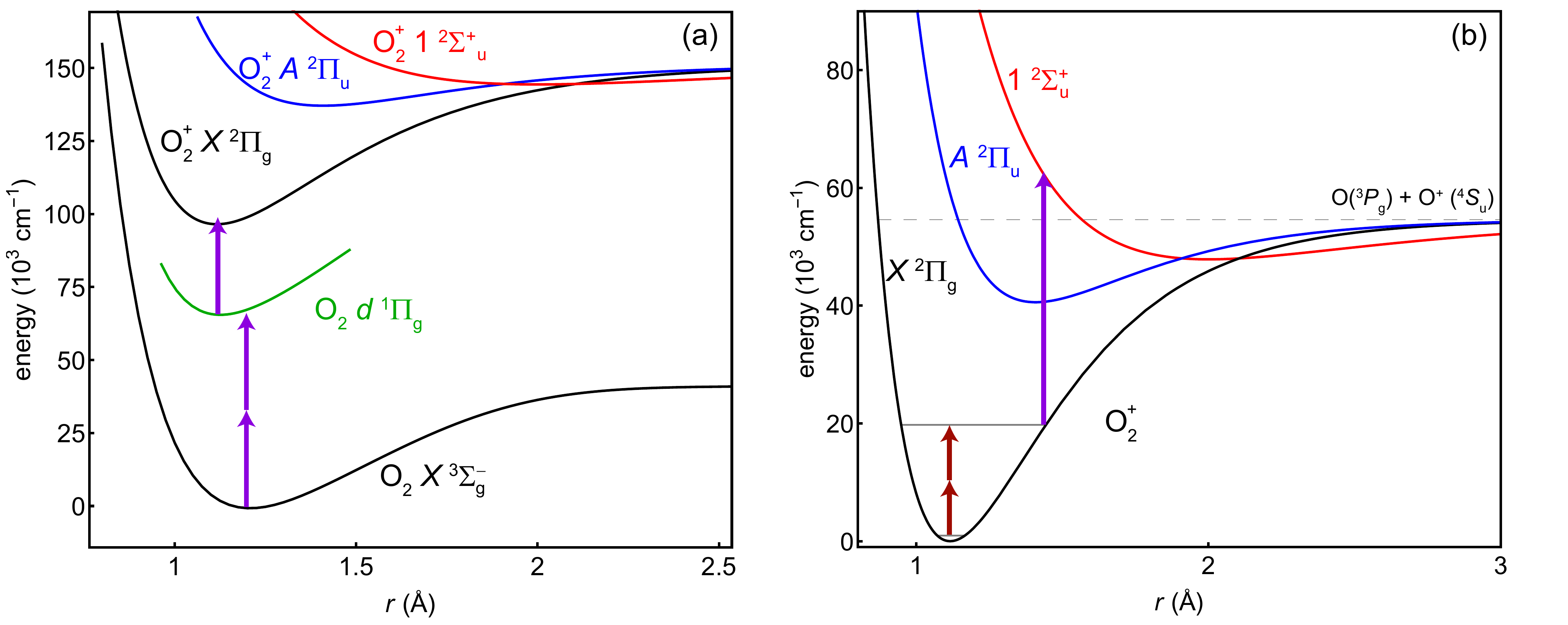}%
\caption{General scheme of the proposed experiments. (a) The \OOp~ion is prepared in its ground vibrational state by photoionization through the neutral \dstate~Rydberg state with (2+1) photons at 301~nm. (b) Two photons drive the vibrational overtone, which is the spectroscopy transition. One photon dissociates any molecules in the excited vibrational state. The overtone shown is $v_X = 11$.}%
\label{fig:scheme}%
\end{figure}

Here, we focus on spectroscopy procedures that are not limited in the number of ions probed. They are general enough to be used in a beam or trap. A trap with co-trapped atomic ions for cooling will be necessary for tests of $\mu$ variation, but initial measurements in a beam could begin narrowing the uncertainties in the transition frequencies. An overview of the procedure is shown in fig.~\ref{fig:scheme}. It takes place in three parts: (1) photoionization assisted by a resonance with a Rydberg state in the neutral O$_2$ (fig.~\ref{fig:scheme}(a)), (2) the two-photon vibrational transition (fig.~\ref{fig:scheme}(b)), (3) selective dissociation of the excited vibrational state followed by measurement of any remaining ions' masses (fig.~\ref{fig:scheme}(b)). The plotted example is for $v_X = 11$, but the techniques are general to many $v_X$.

\subsection{State preparation: Resonance-enhanced multi-photon ionization (REMPI)}

The \Xstate~electronic ground state of \OOp~is of Hund's case (a) and has two fine-structure manifolds with electronic angular momentum quantum numbers $\Omega = \tfrac{1}{2}$ and $\tfrac{3}{2}$. For reasons discussed in section~\ref{sec:othersystematics}, systematic effects are more favorable in the $J=\tfrac{1}{2}$ rotational state~\cite{kajitaPRA2017}. Thus, the initial state for any experiment should be $\ket{\XstateHalf, v=0, J=\tfrac{1}{2}}$.

Much of the data about the \OOp~molecule has been obtained through single-photon photoionization~\cite{kongCJP1994,merktMolPhys1998,songJCP1999,songJCP2000,songJCP2000a} while observing the kinetic energy of the photoelectron. The required vacuum or extreme ultraviolet radiation has been generated from four-wave mixing in gases~\cite{kongCJP1994,merktMolPhys1998} or a synchrotron light source~\cite{songJCP1999,songJCP2000,songJCP2000a}. While ionization into the \OOp~molecule's $\ket{\XstateHalf, v=0, J=\tfrac{1}{2}}$ ground state with one photon would be possible, the 103-nm wavelength is challenging to produce. In general, one-photon ionization into excited rovibrational states would not be possible because of the poor Franck-Condon overlap of the neutral and ionic electronic ground states.

Instead, one can use a bound-to-bound resonance~\cite{prattRPP1995} in neutral O$_2$ with a more conventional laser source to reduce the number of resulting ionic states. In particular, Rydberg excited states can have very high Franck-Condon overlap with the ionic ground state~\cite{morrillJCP1999}. In \OOp, such resonance-enhanced multi-photon ionization (REMPI) schemes have achieved near perfect vibrational selectivity~\cite{surJCP1985,parkJCP1988,looJCp1989,surJCP1991,dochainEPJConf2015}. With a single laser, one can excite a single rotational level in the neutral's Rydberg state, but the ion's rotational distributions will be governed by angular momentum propensities for the bound-to-continuum transition. The addition of a second laser can enhance ionization into the rotational ground state, as has been demonstrated in the N$_2^+$ molecule~\cite{tongPRL2010,tongPRA2011}.

The $[\OOp\,X\,^2\Pi_g]~3s\sigma_g~d\,^1\Pi_g$ Rydberg state provides a suitable resonant state for photoionization. It lies \SI{66380.15}{\per\centi\meter} above the neutral O$_2\,X\,^3\Sigma_g^-$ ground state~\cite{surJCP1991}, such that it can be excited by two 301~nm photons. We use a frequency doubled pulsed dye laser (the dye is a mix of Rhodamine 610 \& 640) to drive the transition. The transition has a linewidth of approximately 2~cm$^{-1}$~\cite{surJCP1991,morrillJCP1999}. 

\begin{figure}%
\includegraphics[width=\columnwidth]{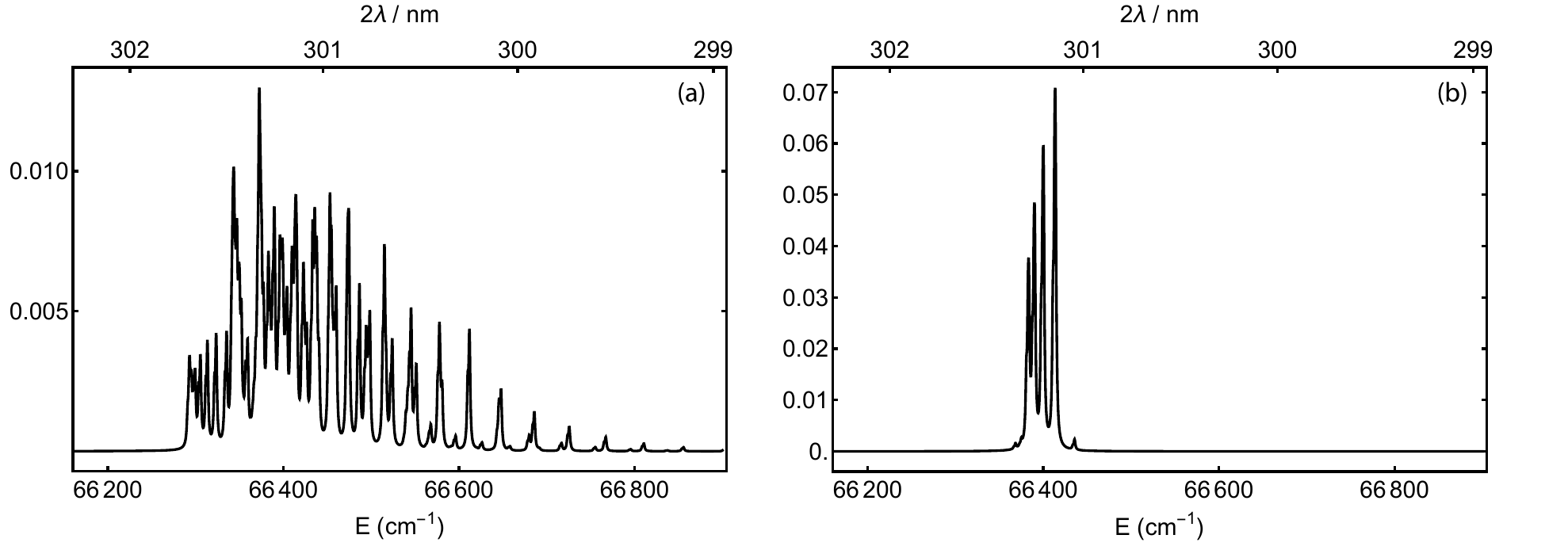}%
\caption{Calculated two-photon excitation spectra from the neutral O$_2$ $X\,^3\Sigma_g^-$ ground state to the $d\,^1\Pi_g$ Rydberg state, both in their ground vibrational state. The temperatures shown correspond to (a) an effusive beam at 300~K or (b) a supersonically expanded one at 5~K. The traces are normalized such that the sum of all transitions is one; note the resulting scale change.}%
\label{fig:dStateSpectrum}%
\end{figure}

Fig.~\ref{fig:dStateSpectrum} shows calculated excitation spectra for the $\textrm{O}_2\,X\rightarrow\rightarrow d$ transition, where the molecules remain in the ground vibrational state. The excitation has seven branches with $\Delta N = 0, \pm1, \pm2, \pm3$. Ref.~\cite{herzbergVolI1950}, eq.~V17, provides the fine-structure splitting and molecular constants for the triplet state. The $^{16}$O nuclei are spin-0 such that only symmetric rotational states exist. For the $^3\Sigma_g^-$ ground state, these are the rotational states with odd-$N$~\cite[Sec.\,V2c]{herzbergVolI1950}.  Refs.~\cite{surJCP1991,morrillJCP1999} provide the $d$-state constants. The figure includes two thermal initial rotational distributions~\cite{herzbergVolI1950}. Fig.~\ref{fig:dStateSpectrum}(a) is calculated at 300~K, which corresponds to an effusive beam or background gas. Because of the finite $d$-state linewidth, the peaks overlap such that a single laser frequency could excite to several different rotational states in $d$. Fig.~\ref{fig:dStateSpectrum}(b) is calculated at 5~K, which is an approximate temperature for a supersonically expanded beam~\cite{surJCP1991,scolesBook1988}. The lower temperature increases the population of the lower rotational states. Most of the remaining overlap comes from the triplet splitting of the ground state rather than any rotational difference in the Rydberg state.

The \OOp ground state is \SI{31070}{\per\centi\meter} above the $\textrm{O}_2\,d\,^1\Pi_g$ state, such that a third photon from the 301-nm laser has sufficient energy to remove an electron. Although it also has enough energy to reach the first vibrationally excited state of \OOp, the near diagonal Franck-Condon factors between the Rydberg and ionic states largely suppress any vibrational excitation. Ionization into the $J=\tfrac{1}{2}$ rotational state could be enhanced with a second laser of wavelength 323~nm, which would need to be tuned to have enough energy to ionize the molecule but not enough to reach the $\ket{\XstateHalf, J=\tfrac{3}{2}}$ state. %

With only a single laser, the ionic rotational states are constrained by symmetry and angular momentum conservation~\cite{xieJCP1990,germannJCP2016}. Any change in the molecule's angular momentum must come from one of three sources: the photoelectron with partial-wave orbital angular momentum $l$, the photoelectron's spin of $\frac{1}{2}$, or one unit of orbital angular momentum from the photon. The $g\leftrightarrow g$ nature of the transition constrains $l$ to be odd. The net orbital angular momentum transferred to the molecule can take values $k = l-1$ or $l+1$. Including the photoelectron's spin, the total angular momentum transferred to the molecule can take values $u=|k-\tfrac{1}{2}|$ or $k+\tfrac{1}{2}$, such that the total angular momentum can change by $|\Delta J|\le u$. For example, orbital angular momentum transfers of $k = 0$ or 2 would allow
\begin{subequations}
	\begin{align}
	\Delta J &= \pm\tfrac{1}{2} &  (k&=0) \\
	\Delta J &= \pm\tfrac{1}{2},~\pm\tfrac{3}{2},~\pm \tfrac{5}{2} & (k&=2).
	\end{align}
\end{subequations}
 In general, many photoionization processes show a higher propensity for transitions with $k=0$. Transitions with $k=2$ are somewhat less likely, and those with $k>2$ contribute very little. To say more about their relative probabilities in the ${\rm O}_2\,\dstate\rightarrow\OOp\,\Xstate$ transition would require either calculations with a model of the molecules' orbitals or measurements of the relative strengths of each rotational transition.

The four prominent peaks in the low-temperature distribution of fig.~\ref{fig:dStateSpectrum}(b) are excitations from $\ket{{\rm O}_2\,X\,^3\Sigma_g^-, v=0, N=1}$ to $\ket{{\rm O}_2\,\dstate, v^\prime=0, N^\prime = 1, 2, 3, 4}$, in order from lowest to highest energy. The $N^\prime=1,2$ states can ionize into our target $\ket{\OOp\,\XstateHalf, v=0, J = 1/2}$ state with a $k=0$ transition. The $N^\prime = 3$ state would require $k=2$. The resonance for $N^\prime=1$ is \SI{66383.5}{\per\centi\meter}, which is two $301.280~{\rm nm}$ photons. For $N^\prime=2$ it is \SI{66390.2}{\per\centi\meter}, which is two $301.249~{\rm nm}$ photons. In both cases, the peak is three overlapping transitions from the neutral's triplet ground state and is broader to the low-energy side.

Photoionization is a useful state preparation technique for any number of molecular ions. With a single photoionization laser, some molecules will be in the wrong $J$ state. For multiple-ion experiments, this simply eats into the signal. For single-ion experiments, one could reload the molecule until the $J=\tfrac{1}{2}$ state is achieved or could manipulate the rotational states with a quantum-projection scheme. The 50~GHz rotation constant~\cite{coxonJMS1984} means $J=\tfrac{3}{2}$ and $\tfrac{1}{2}$ are 151~GHz away, so a broadband/femtosecond laser would be required~\cite{leibfriedNJP2012,dingNJP2012}.

\subsection{Probe: Two-photon transition}

The \OOp~\Xstate~potential is \SI{54600}{\per\centi\meter} deep with approximately 55 vibrational levels~\cite{songJCP1999}. In principle, two-photon transitions can be driven from $v=0$ to any $v_X$. The transition frequency (for example in hertz) for the $v = 0\rightarrow\rightarrow v_X$ vibrational overtone is
\begin{equation}
	f = \left(E_{\ket{X, v_X}} - E_{\ket{X, v = 0}}\right)/h.
	\label{eq:transitionFreq}
\end{equation}
Analysis of emission spectra~\cite{coxonJMS1984} give values for $E_{\ket{X, v_X}}$, including high-precision fine-structure and rotational constants, for $v_X = 0-11$. Pulsed-field-ionization photoelectron (PFI-PE) studies~\cite{songJCP1999} give vibrational, fine-structure, and rotational constants up to $v_X = 38$. Uncertainties on the vibrational energies range from 1--150~GHz. Since all electric-dipole transitions within \Xstate~are forbidden, the two-photon transition is intrinsically narrow, with linewidths limited by the probe laser.

The laser wavelengths required to drive the overtones range from \SI{10.6}{\micro\meter} ($v_X = 1$) to \SI{418}{\nano\meter} ($v_X = 38$) and beyond. Table~\ref{tab:2E1} provides a list. The table also calculates the sensitivity to $\mu$-variation $f_\mu$, which grows linearly for low $v_X$ and peaks at $v_X = 28$.

Searches for drifts in $\mu$ consist of monitoring the transition frequency multiple times for a reasonable duration such as a year. Because the probe laser's coherence time will be much less than a year, a second signal with different $f_\mu$ will be required, such as from an optical atomic clock or a second transition in \OOp (see section~\ref{sec:reference}). Fourier analysis of the measurements can probe for low-frequency oscillations in $\mu$. Sensitivity at higher frequencies can be enhanced by use of a composite pulse sequence, which are usually designed to suppress noise but can also be designed to enhance signals at particular frequencies.

\subsection{Detection: Selective dissociation}

Detection of a successful transition to $v_X$ could be done by selective dissociation of that state~\cite{bertelsenJPB2006,rothPRA2006,seckJMS2014,niJMS2014} 
 or energy-dependent chemical reaction such as charge transfer with a background gas~\cite{schlemmerIJMS1999,tongPRL2010}, followed by observation of ion loss or the mass of any remaining ions. These are destructive and thus suitable to experiments with large numbers of ions, where the increased statistics justifies the experiment down time for reloading. For few- or single-ion experiments, as will be required at the highest precision, the more complex quantum-logic techniques~\cite{schmidtScience2005,mur-petitPRA2012,wolfNature2016,chouNature2017} will be preferable.

The \OOp~molecule has a relatively shallow bound \SigmaState~state~\cite{coxonJMS1984,akahoriJCP1997,fedorovJCP2001,liuMolPhys2015} that appears suitable for dissociation to $\textrm{O}(^3P_g)+\textrm{O}^+(^4S_u)$. Its potential minimum is offset relative to \Xstate~such that many vibrational states $v_X$ have good overlap with dissociating states.

Table~\ref{tab:2E1} has a list of preferred dissociation laser wavelengths $\lambda_{\rm PD}$ as a function of $v_X$. This somewhat naive %
list simply finds the energy difference between the outer turning point of the \Xstate~potential and the inner turning point of \SigmaState. For the $X$ potential, we use our own calculation (see sec.~\ref{sec:dipoleCalc}); for the \SigmaState~potential, we use a Morse approximation~\cite{herzbergVolI1950} based on coefficients in ref.~\cite{liuMolPhys2015}. Photodissociation cross-sections are generally not so sensitive to wavelength, so other wavelengths should work as well. Fig.~\ref{fig:scheme}(b) shows an example of the 238~nm dissociation of $v_X=11$. For states with $v_X\ge17$, the inner turning point of \SigmaState~corresponds to a bound state, so a more in-depth look at photodissociation cross-sections -- or a different detection scheme or dissociating state -- is warranted.

The presence of atomic oxygen ions would thus signify a successful two-photon probe transition. The factor of two mass difference between $\textrm{O}^+$ and \OOp~should be easy to resolve. In a radiofrequency trap, the charge-to-mass ratio of the ions can be determined by their spatial distribution~\cite{zhangPRA2007,tongPRL2010} (lighter ions are trapped more deeply) or by resonantly exciting the mass-dependent radial trap frequency~\cite{babaJJAP1996,babaJAP2002,schneiderNPhys2010}. Alternatively, ions in free space (either a beam or a rapidly quenched trap) can be accelerated into a time-of-flight mass spectrometer~\cite{schowalterRSI2012,seckJMS2014,debPRA2015,schneiderIJMS2016,schmidRSI2017}. Since the dissociation is itself destructive, the loss of the ions in the time-of-flight technique is not a concern. It is also fast and has high signal-to-noise.

\section{Transition rates and electric-dipole-related systematics} \label{sec:calculations}

The two-photon spectroscopy transition is enabled by electric-dipole coupling to excited electronic states. This coupling also creates a mechanism for systematic shifts from the driving laser, trapping field, and blackbody radiation. We estimate these rates and shifts by assuming the coupling is primarily through the \Astate~state, which is the lowest-energy state with the same spin multiplicity as the \Xstate~state. The large detuning of this state (the lowest vibrational states of $X$ and $A$ are \SI{40070}{\per\centi\meter} = 1.20~PHz/$c$ apart~\cite{coxonJMS1984,songJCP1999,songJCP2000}) suppresses these systematic effects but also increases the laser intensity required for the two-photon transition. We do not include other excited states because they have either a much larger detuning, much smaller Franck-Condon overlap (for example the \SigmaState~state), a different spin multiplicity (for example the $a\,^4\Pi_u$ state), or the wrong center-of-inversion symmetry (g/u). We also neglect continuum transitions.

\subsection{Calculating the transition dipole moments}\label{sec:dipoleCalc}

The perturbation-theory calculations for transition rates and systematic frequency shifts require electric dipole moments and energy differences between $X$ and $A$ vibrational states. The transition dipole moment between the states $\ket{X, v_X}$ and $\ket{A, v_A}$ is given by
\begin{equation}
	{D}_{v_A v_X} = \braOket{A, v_A}{{d}}{X, v_X},
\end{equation}
where ${d}$ is the sum of all the charges' electric dipole moments. 
We ignore rotation and fine-structure in this calculation. These splittings are of order \SI{1}{\per\centi\meter} for vibrational levels in the \Astate~state~\cite{coxonJMS1984,songJCP2000} such that any change in the detuning is negligible. The calculations below involve sums over states, and the sum over the rotational and fine-structure states is effectively the identity.

We can separate the electronic and vibrational contributions~\cite{herzbergVolI1950,LefebvreBrionAndField} 
and write the transition dipole moment in terms of the vibrational wavefunctions, internuclear distance $r$, and the electronic transition dipole moment $D_e = \braOket{A}{d}{X}$:
\begin{equation}
	D_{v_A v_X} = \int \psi^*_{v_A}\!(r)\,D_e(r)\,\psi_{v_X}\!(r)\,dr.
\end{equation}
In general, $D_e$ is a function of the internuclear distance. In most cases, including here, this function varies slowly, and we may replace it with a constant evaluated at the $r$-centroid~\cite{herzbergVolI1950,LefebvreBrionAndField,gilmoreJPCRD1992}
\begin{equation}
	\bar{r}_{v_A v_X} = \int \psi^*_{v_A}\,r\,\psi_{v_X}\,dr / \int \psi_{v_A}^* \psi_{v_X}\,dr.
\end{equation}
The resulting transition dipole moment is given by
\begin{equation}
	D_{v_A v_X} = D_e(\bar{r}_{v_A v_X}) \int \psi^*_{v_A}\psi_{v_X}\,dr.
	\label{eq:dipolemoment}
\end{equation}
The wavefunction-overlap integral in eq.~\ref{eq:dipolemoment} is the square-root of the Franck-Condon factor. Importantly -- and unlike the Franck-Condon factor -- it is a signed quantity that can interfere in a sum over vibrational states.

Calculated tables of $X$--to--$A$ transition dipole moments and energies exist~\cite{gilmoreJPCRD1992}, however they only tabulate values up to the 21$^{\rm st}$ vibrational state in each potential well. For some two-photon transitions, we found that contributions from higher-lying vibrational levels in the $A$ potential were needed. We therefore calculated the terms ourselves.

As no recent experimental potentials were available, we performed our own RKR calculation using the %
 RKR1 2.0 program~\cite{leroyJQSRT2017} with data from refs.~\cite{songJCP1999,songJCP2000}. The \Xstate~state RKR, using the first 39 levels, was performed using a Dunham expansion to 19 terms in $G_v$ and $B_v$. The \Astate~state RKR, using the first 13 levels, was performed with a Dunham expansion to 11 terms in $G_v$ and 7 terms in $B_v$. The resulting potential for the $A$ state was extended to long range by fitting for $C_6$, $C_8$, and $C_{10}$ coefficients using the last 5 RKR points. Using these potentials, the vibrational state energies and wavefunctions were generated using the program LEVEL 8.2%
~\cite{leroyJQSRT2017a} with 0.001 angstrom resolution. Root-mean-square residuals for the energies were \SI{4}{\per\centi\meter} for $X$ and \SI{3}{\per\centi\meter} for $A$; all residuals were less than \SI{10}{\per\centi\meter}. We then numerically calculated the wavefunction overlap integrals and $r$-centroids. We calculated the electronic transition dipole moment in the $r$-centroid approximation $D_e(\bar{r}_{v_A v_X})$ using the fit to theoretical values of $D_e(r)$ found in ref.~\cite{gilmoreJPCRD1992}.

\begin{figure}%
\includegraphics[width=\columnwidth]{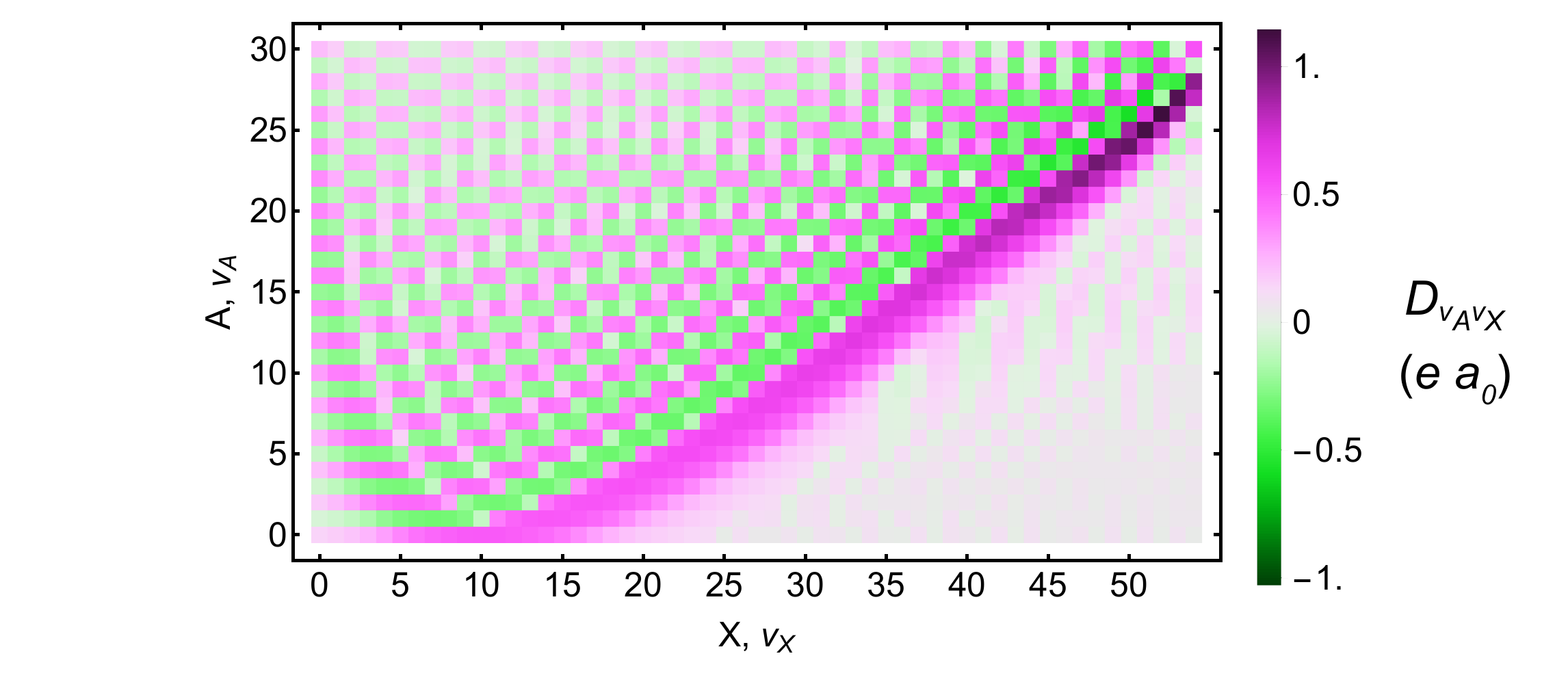}%
\caption{Calculated transition dipole moments between \Xstate~and \Astate~vibrational states. Both the electronic and vibrational contributions are included. Dipole moments are given in atomic units. Numeric values are available in the supplementary material.}%
\label{fig:dipoleMoments}%
\end{figure}

Fig.~\ref{fig:dipoleMoments} shows the transition dipole moment between each $X$ and $A$ vibrational state. In the supplementary material accompanying this manuscript, we provide our RKR potentials; wavefunctions, energies, and rotational constants for 55 vibrational states within $X$ and 31 states within $A$; as well as overlap integrals, energy differences, and transition dipole moments for all pairs of vibrational states.

\begin{table}
	\caption{Table of two-photon-transition $(v=0\rightarrow\rightarrow v_X)$ information: the wavelength required to drive the transition, the absolute sensitivity of the transition to $\mu$ variation, the optimal photodissociation wavelength, and the calculated results for transition rates and electric-dipole-related systematics. These results are listed as the two-photon Rabi frequency (in terms of intensity), the probe laser AC Stark shift (in terms of intensity and of Rabi frequency), the AC Stark shift from the trapping field (in terms of the mean-square electric field), and the blackbody radiation shift at 300~K. The frequency shifts are normalized by the magnitude of the absolute sensitivity $f_\mu$ so the numbers represent relative accuracy in $\Delta\mu/\mu$ and the signs reflect the actual frequency shift.} 	\label{tab:2E1}
	\centering
\resizebox{\columnwidth}{!}{ %
	\begin{tabular}{crrcSSSSS} %
	\toprule
	$\mathbf{v_X}$  & 
	\makecell{$\boldsymbol{2\lambda}$ \\ \textbf{(nm)}} & 
	\makecell{$\mathbf{f}\boldsymbol{_\mu}$ \\ \textbf{(THz)}} & 
	\makecell{$\boldsymbol{\lambda}_\textbf{PD}$ \\ \textbf{(nm)}} &
	{\makecell{$\frac{\boldsymbol{\Omega}_\mathbf{R}}{\mathbf{2}\boldsymbol{\pi}}\mathbf{ / I }$\\$\mathbf{(10^{-7} \frac{\bf Hz}{\bf{W/m}^2})}$} } &
	{\makecell{$\frac{\boldsymbol{\Delta}\mathbf{f}_{\bf probe}}{\textbf{|}\mathbf{f}\boldsymbol{_\mu}\textbf{|}}\mathbf{ / I }$ \\ $\textbf{(}\mathbf{10^{-20}}\,\textbf{(W/m}\mathbf{^2}\textbf{)}\mathbf{^{-1}}\textbf{)}$} } &
	{\makecell{$\frac{\boldsymbol{\Delta}\mathbf{f}_{\bf probe}}{\textbf{|}\mathbf{f}\boldsymbol{_\mu}\textbf{|}}\mathbf{ / }\frac{\boldsymbol{\Omega}_\mathbf{R}}{\mathbf{2}\boldsymbol{\pi}}$ \\ $\textbf{(}\mathbf{10^{-13}}\,\textbf{(Hz)}\mathbf{^{-1}}\textbf{)}$} }&
	{\makecell{$\frac{\boldsymbol{\Delta}\mathbf{f}_{\bf trap}}{\textbf{|}\mathbf{f}\boldsymbol{_\mu}\textbf{|}}\mathbf{ / \mathcal{E}}_\textbf{rms}^\mathbf{2}$ \\ $\textbf{(}\mathbf{10^{-22}}\,\textbf{(V/m)}\mathbf{^{2}}\textbf{)}$} } &
	{\makecell{$\frac{\boldsymbol{\Delta}\mathbf{f}_{\bf BBR}}{\textbf{|}\mathbf{f}\boldsymbol{_\mu}\textbf{|}}$ \\ $\textbf{(}\mathbf{10^{-18}}\textbf{)}$} } \\
	\midrule
 1 & 10614 & -28 & 113 & 4.94 & -0.59 & -0.12 & -0.16 & -3.59 \\
 2 & 5386 & -54 & 123 & 1.24 & -0.40 & -0.32 & -0.11 & -2.43 \\
 3 & 3617 & -80 & 134 & 1.56 & -0.47 & -0.30 & -0.12 & -2.86 \\
 4 & 2738 & -104 & 144 & 0.91 & -0.45 & -0.49 & -0.12 & -2.69 \\
 5 & 2211 & -128 & 155 & 0.99 & -0.51 & -0.52 & -0.13 & -3.07 \\
 6 & 1859 & -151 & 166 & 0.91 & -0.53 & -0.59 & -0.14 & -3.14 \\
 7 & 1609 & -172 & 179 & 0.68 & -0.62 & -0.90 & -0.15 & -3.56 \\
 8 & 1421 & -193 & 192 & 0.91 & -0.67 & -0.73 & -0.16 & -3.79 \\
 9 & 1275 & -213 & 206 & 0.51 & -0.78 & -1.53 & -0.18 & -4.26 \\
 10 & 1158 & -231 & 221 & 0.92 & -0.88 & -0.95 & -0.20 & -4.59 \\
 11 & 1063 & -249 & 238 & 0.40 & -1.03 & -2.56 & -0.22 & -5.14 \\
 12 & 984 & -266 & 256 & 0.93 & -1.22 & -1.31 & -0.24 & -5.65 \\
 13 & 917 & -281 & 276 & 0.33 & -1.53 & -4.57 & -0.28 & -6.39 \\
 14 & 860 & -296 & 299 & 0.94 & -1.97 & -2.09 & -0.31 & -7.14 \\
 15 & 810 & -310 & 323 & 0.30 & -2.79 & -9.20 & -0.35 & -8.14 \\
 16 & 767 & -323 & 351 & 0.96 & -4.72 & -4.91 & -0.40 & -9.20 \\
 17 & 730 & -334 & - & 0.31 & -1.48 & -4.77 & -0.46 & -10.58 \\
 18 & 696 & -345 & - & 0.98 & 9.10 & 9.29 & -0.52 & -12.06 \\
 19 & 667 & -355 & - & 0.35 & 17.52 & 49.40 & -0.60 & -13.95 \\
 20 & 640 & -364 & - & 1.01 & 1.79 & 1.77 & -0.69 & -16.06 \\
 21 & 616 & -371 & - & 0.43 & -6.69 & -15.59 & -0.81 & -18.81 \\
 22 & 594 & -378 & - & 1.01 & 5.88 & 5.79 & -0.95 & -21.97 \\
 23 & 575 & -384 & - & 0.56 & 3.55 & 6.30 & -1.12 & -26.10 \\
 24 & 557 & -389 & - & 1.01 & 3.34 & 3.29 & -1.32 & -30.82 \\
 25 & 541 & -393 & - & 0.75 & 2.66 & 3.55 & -1.58 & -36.94 \\
 26 & 526 & -395 & - & 0.99 & 2.18 & 2.21 & -1.88 & -43.99 \\
 27 & 513 & -397 & - & 0.99 & 2.04 & 2.06 & -2.27 & -53.31 \\
 28 & 500 & -398 & - & 0.86 & 1.85 & 2.14 & -2.74 & -64.36 \\
 29 & 489 & -398 & - & 1.28 & 1.80 & 1.40 & -3.35 & -79.02 \\
 30 & 478 & -397 & - & 0.59 & 1.71 & 2.88 & -4.09 & -96.75 \\
 31 & 468 & -395 & - & 1.58 & 1.69 & 1.07 & -5.06 & -120.31 \\
 32 & 459 & -392 & - & 0.22 & 1.66 & 7.62 & -6.29 & -150.64 \\
 33 & 451 & -388 & - & 1.84 & 1.68 & 0.91 & -7.92 & -191.28 \\
 34 & 444 & -383 & - & 0.38 & 1.72 & 4.47 & -9.98 & -242.92 \\
 35 & 437 & -377 & - & 1.91 & 1.73 & 0.91 & -12.32 & -303.11 \\
 36 & 430 & -369 & - & 1.21 & 1.78 & 1.47 & -15.20 & -379.12 \\
 37 & 424 & -361 & - & 1.69 & 1.77 & 1.05 & -18.99 & -483.29 \\
 38 & 418 & -352 & - & 2.28 & 1.86 & 0.82 & -24.72 & -646.53 \\
\bottomrule
\end{tabular}
} %

\end{table}

\subsection{Transition rate}

The two-photon transition rate for the $\ket{X, v=0}\rightarrow\rightarrow\ket{X, v_X}$ transition, stated as a Rabi frequency is%
\begin{equation}
	\frac{\Omega_R}{2\pi} = \frac{I}{\epsilon_0 c h^2}\sum_{v_A = 0}^{\maxA}\frac{D_{v_A v_X} D_{v_A 0}}{f_{v_A 0}-f_L}.
\end{equation}
Here, 
\begin{equation}
	f_{v_A v_X} = \left(E_{\ket{A, v_A}} - E_{\ket{X, v_X}}\right)/h
\end{equation}
is the transition frequency (for example in hertz) between the $A$ and $X$ vibrational states, $f_L$ is the laser frequency, which on resonance is $f/2$ or half the vibrational overtone frequency, and $I$ is the laser intensity. The Rabi frequency $\Omega_R$ is an angular frequency, but we state our results in terms of the actual frequency $\Omega_R/(2\pi)$.

The signs of the dipole moments are important in the sum and cause the terms to interfere with each other. Some other references, such as ref.~\cite{gilmoreJPCRD1992}, do not include the signs in their tables.

Table~\ref{tab:2E1} lists the Rabi frequency in terms of laser intensity. It includes all two-photon transitions up to $v_X=38$, which is the highest level for which spectroscopic data is available. Note that the $a\,^4\Pi_u$ potential overlaps with $v_X\ge 21$ and the \Astate~potential overlaps with $v_X\ge 28$. Because the probe photons are only half the transition energy, they remain far detuned. After accounting for fine-structure and rotational splittings, there may be some near-degeneracies among the $X$, $A$, and $a$ states. These could provide alternate routes to $\mu$-variation measurements~\cite{hannekePRA2016}. For the purposes of the two-photon experiments, they could cause the eigenstates to mix such that new systematic effects creep in. Because both $A$ and $a$ states have ungerade symmetry, they should not mix with the gerade $X$ state.

To achieve a reasonable transition probability requires either high intensity or long interrogation time. For example, a 1~W laser focused to a \SI{150}{\micro\meter} waist has an intensity of $2.8\times10^7~{\rm W}/{\rm m}^2$. When resonant with the $\ket{X, v=0}\rightarrow\rightarrow\ket{X, v_X=11}$ transition, that laser would produce a Rabi frequency of 1~Hz.

\subsection{Stark shifts}

AC Stark shifts will be created by oscillating electric fields. Three main sources in this experiment are the probe laser itself, the trapping field of a radiofrequency trap, and blackbody radiation. An electric field of with root-mean-square (rms) value $\mathcal{E}_{\rm rms}$ oscillating at frequency $f_L$ will shift the frequency of the state $\ket{X, v_X}$ due to interactions with all vibrational states in $\Astate$:
\begin{equation}
	\delta f_{v_x} = -\frac{\mathcal{E}_{\rm rms}^2}{h^2} \sum_{~~v_A=0}^{\maxA} D_{v_A v_X}^2 \frac{f_{v_A v_X}}{f_{v_A v_X}^2 - f_L^2}.
	\label{eq:ACStark}
\end{equation}

The net shift on the $\ket{X, v=0}\rightarrow\rightarrow\ket{X, v_X}$ overtone transition frequency is
\begin{equation}
	\Delta f = \delta f_{v_X} - \delta f_0.
		\label{eq:netACStark}
\end{equation}

\subsubsection{Probe laser $(\Delta f_{\rm probe})$}

For the probe laser, it is more useful to talk in terms of the intensity
\begin{equation}
	I = c\epsilon_0 \mathcal{E}_{\rm rms}^2.
	\label{eq:intensity}
\end{equation}
Both the Rabi frequency and AC Stark shift are proportional to the laser's intensity, such that simple Rabi-style probes will have their Stark shifts increasing linearly with transition rate. Table~\ref{tab:2E1} lists the net AC Stark shift of the probe laser $\Delta f_{\rm probe}$ in terms of both the laser intensity and the transition Rabi frequency. In each case, we normalize the result by the absolute sensitivity $f_\mu$ (see eq.~\ref{eq:absSens}) to show how it might affect a $\mu$-variation measurement. We use the magnitude of $f_\mu$ to emphasize the sign of the shift on the actual frequency; the shift on $\Delta\mu/\mu$ is the opposite sign.

The above example of a 1~W laser focused to \SI{150}{\micro\meter} would produce a \SI{-70}{\hertz} AC Stark shift on the $\ket{X, v=0}\rightarrow\rightarrow\ket{X, v_X=11}$ transition. This is a fractional shift of $-1\times10^{-13}$ in frequency and $3\times10^{-13}$ in $\Delta\mu/\mu$. Thus the probe laser's AC Stark shift has the potential to be a major systematic effect. For lower-precision experiments with stable laser intensity, the Stark shift could be calibrated. Better would be to use a composite-pulse scheme such as hyper- and autobalanced Ramsey spectroscopy~\cite{yudinPRA2010,yudinPRAppl2018,sannerPRL2018,zanonwilletteRPP2018}. Such schemes have suppressed AC Stark shifts by four orders of magnitude when used on the electric octupole ($E3$) clock transition in {Yb$^+$}~\cite{huntemannPRL2012,huntemannPRL2016,sannerPRL2018}.

\subsubsection{Trapping fields $(\Delta f_{\rm trap})$}

Experiments with longer probe-times will require the ions to be held in place. In a radiofrequency (rf) trap, the trapping field itself can produce an AC Stark shift. Since the rf trapping fields are much lower in frequency than any $f_{v_A v_X}$, we calculate the shift $\Delta f_{\rm trap}$ in the limit $f_L\rightarrow 0$, and tabulate the results in terms of $\mathcal{E}_{\rm rms}^2$ in table~\ref{tab:2E1}.

For a linear rf trap, the field is established by electrodes $R$ away from the ions and oscillating at potentials $V_0\cos(\Omega t)$. These produce an electric field~\cite{berkelandJAP1998}
\begin{equation}
	\boldsymbol{\mathcal{E}}_{\rm rms} = -\frac{V_0}{\sqrt{2}R^2}(x \,\mathbf{\hat{x}}-y\,\mathbf{\hat{y}}) = -\frac{m\Omega^2}{2\sqrt{2}Q}(q_x x \,\mathbf{\hat{x}} + q_y y \,\mathbf{\hat{y}}),
	\label{eq:trapfield}
\end{equation}
where $m/Q$ is the mass-to-charge ratio of the trapped ion, and $q_{x,y}$ are parameters in a Mathieu equation describing the ions' motion (see for example~\cite{berkelandJAP1998,winelandJRNIST1998}). These trapping parameters can be calibrated in place by use of the motional frequencies of the trapped ions. Typical values in our apparatus are $q_x = -q_y \approx 0.1$ 
and $\Omega \approx 2\pi(10~\textrm{MHz})$ such that \OOp~molecular ions would experience a field curvature of  $5\times10^7~\textrm{V/m}^2$. Ions displaced \SI{100}{\micro\meter} off the rf null -- as might happen in a 3D crystal -- would experience an electric field of $\mathcal{E}_{\rm rms} \sim 5\times10^3~\textrm{V/m}$. For the $\ket{X, v=0}\rightarrow\rightarrow\ket{X, v_X=11}$ transition, this would lead to a shift of \SI{-0.1}{\hertz}, which is a fractional shift of $-2\times10^{-16}$ in frequency and $5\times10^{-16}$ in $\Delta\mu/\mu$. As shown in section~\ref{sec:Doppler}, displacement off the rf null also produces a Doppler shift that is more of a concern.

\subsubsection{Blackbody radiation $(\Delta f_{\rm BBR})$}

The blackbody radiation (BBR) from the environment causes far off-resonant AC Stark shifts. To determine the shift of a given $X$-state vibrational level, we integrate over the continuous blackbody spectrum. At temperature $T$, the shift of level $v_X$ is given by~\cite{farleyPRA1981,porsevPRA2006}
\begin{equation}
	\delta f_{v_X} = -\frac{4\pi}{3\epsilon_0 h c^3}\left(\frac{kT}{h}\right)^{\!\!3}~~\sum_{v_A = 0}^{\maxA} D_{v_A v_X}^2 ~F\!\left[\frac{h f_{v_A v_X}}{kT}\right],
\end{equation}
where
\begin{equation}
	F[y] = \int_0^\infty \left(\frac{1}{y+x}+\frac{1}{y-x}\right) \frac{x^3\,dx}{e^x-1}.
\end{equation}
The $1/(y-x)$ part of the integral is evaluated with the Cauchy principal value at the pole~\cite{farleyPRA1981}.

The net shift $\Delta f_{\rm BBR}$ is calculated at 300~K and included in table~\ref{tab:2E1}. At 300~K, the blackbody shift of the $\ket{X, v=0}\rightarrow\rightarrow\ket{X, v_X=11}$ transition is \SI{-1}{\milli\hertz}, which is a fractional shift of $-2\times10^{-18}$ in frequency and $5\times10^{-18}$ in $\Delta\mu/\mu$. For other temperatures, the overall shift scales as $T^4$~\cite{farleyPRA1981}. For example, at 100~K, the shift would be $3^4=81$ times smaller. For comparison, the Al$^+$ and Lu$^+$ optical clocks~\cite{rosenbandScience2008,chouPRL2010}, which have the smallest blackbody shifts of currently used optical clocks, have room-temperature shifts $-8\times10^{-18}$~\cite{rosenbandArxiv2006} and $-1.4\times10^{-18}$~\cite{arnoldNatureComm2018}. Blackbody radiation will not be a major concern.

\section{Additional systematic effects} \label{sec:othersystematics}

\subsection{Doppler shifts}\label{sec:Doppler}

First-order Doppler shifts are highly suppressed when the ions are trapped~\cite{dickePR1953}. Any possible first-order shift, such as the trap itself moving relative to the laser, can be monitored with counter-propagating probe beams~\cite{rosenbandScience2008}. Second-order Doppler shifts are a relativistic time-dilation effect. They arise from finite temperature and micromotion~\cite{berkelandJAP1998}, which each cause a nonzero mean-square velocity. The shift is equal to
\begin{equation}
\frac{\Delta f}{f} = - \frac{v_{\rm rms}^2}{2c^2}.
\end{equation}
Co-trapped atomic ions can sympathetically cool the molecular motion to reduce any thermal Doppler shifts. For example, at the 0.5~mK theoretical Doppler-cooling limit of \Be, the thermal motion has $v_{\rm rms}^2 \sim kT/m$, such that the second-order Doppler shift would be $\sim -1\times10^{-18}$.

Excess micromotion, however, can produce significant shifts. An ion displaced $x$ off the trap's rf null has a velocity of $v_{\rm rms} = x q_x \Omega/(2\sqrt{2})$~\cite{berkelandJAP1998}, where $q_x$ and $\Omega$ are the same Mathieu trap parameter and angular trap frequency used in Eq.~\ref{eq:trapfield}. This shift could be created, for example, by a static electric field $\mathcal{E}_{\rm DC}$ from other ions or an uncompensated offset in the trap potentials. This gives a second-order Doppler shift~\cite{berkelandJAP1998} of
\begin{equation}
	\frac{\Delta f}{f} = -\frac{x^2 q_x^2 \Omega^2}{16 c^2} = -\frac{4}{m^2 c^2}\left(\frac{Q \mathcal{E}_{\rm DC}}{q_x \Omega}\right)^2.
\end{equation}
For our typical parameters, %
 the fractional frequency shift is $-3\times10^{-17}x^2$~\si{\per\micro\meter\tothe{2}} %
 or $-1\times10^{-17}\mathcal{E}_{\rm DC}^2~({\rm V/m})^{-2}$. An ion in a 3D crystal displaced \SI{100}{\micro\meter} off the rf null would experience a fractional shift of $-3\times10^{-13}$ in frequency and $7\times10^{-13}$ in $\Delta\mu/\mu$. Here again we have the potential for an important systematic effect. Any high-precision work will require ions to be close to the rf null with stray fields well compensated. For comparison, by use of sub-Doppler cooling in a well-compensated trap, an optical clock based on the $^{27}{\rm Al}^+$ ion (comparable in mass to \OOp) has reduced the time-dilation frequency shift to $-(1.9\pm0.1)\times10^{-18}$~\cite{chenPRL2017}.

\subsection{Electric quadrupole shift}

An electric quadrupole shift arises when a non-zero quadrupole electric field interacts with an electric quadrupole moment of the ion~\cite{itanoJRNIST2000,bakalovAPB2014}. For example, Eq.~\ref{eq:trapfield} shows that the ion trap itself produces a nonzero quadrupolar field. For an arbitrary molecular state $\ket{X\,^2\Pi_{g\Omega}\,v_X\,J\,M_J}$, the shift scales as $3M_J^2-J(J+1)$~\cite{itanoJRNIST2000,bakalovAPB2014}. For states with $J=0$ or $\frac{1}{2}$, the quadrupole moment is exactly zero and so the shift is also exactly zero. We thus choose to use the $J=\tfrac{1}{2}$ state in the $\Omega=\tfrac{1}{2}$ manifold of \Xstate.

\subsection{Zeeman shift}

The diagonal matrix element for the Zeeman shift is~\cite{schadeeJQSRT1978,berdyuginaAA2002}
\begin{equation}
	\braOket{\Lambda\,\Sigma;\Omega\,J\,M_J}{-\mathbf{d}_m\cdot\mathbf{B}}{\Lambda\,\Sigma;\Omega\,J\,M_J} = \frac{\mu_B B \Omega}{J(J+1)}(g_S \Sigma+g_L\Lambda)M_J,
\end{equation}
where $B$ is the magnetic field, $\mu_B$ is the Bohr magneton, and $g_S$ and $g_L$ are the spin and orbital $g$-factors. For a $^2\Pi_{\frac{1}{2}}$ state, $\Lambda = -2\Sigma$. Since $g_L=1$ and $g_S$ is only a part-per-thousand larger than 2, the Zeeman shift in that state is suppressed by approximately $10^3$. For $J=\frac{1}{2}$, the shift would be 11~MHz~$M_J$/T (1.1~kHz~$M_J$/G). For transitions between the same $M_J$ ($\pm \frac{1}{2} \leftrightarrow \pm \frac{1}{2}$) the shift cancels. For transitions from $M_J=\frac{1}{2}$ to $-\frac{1}{2}$, the opposite transition could be probed and the two values averaged. 

Higher-order Zeeman shifts from off-diagonal mixing with the $^2\Pi_{g,\frac{3}{2}}$ fine-structure manifold should be largely suppressed because that manifold does not have a $J=\frac{1}{2}$ level. An intrinsic second-order Zeeman shift~\cite{schiffPR1939,garstangRPP1977} arises from the Hamiltonian
\begin{equation}
	H = \frac{q^2}{8 m}(x^2+y^2)B^2,
\end{equation}
where the magnetic field is assumed to be in the $z$-direction. To estimate its size, if $\langle x^2+y^2\rangle$ is of order the Bohr radius squared $a_0^2$, then the shift should be of order $1.5\times10^4$~Hz/T$^2$. A bias field of 1~mT would cause each energy level to shift around 15 mHz. Because the electronic state is unchanged, it is likely that the differential shift in the transition frequency would be much smaller.

Since the $^{16}$O nuclei are spin-0, there are no Zeeman shifts from nuclear spin or hyperfine structure. Zeeman shifts of each energy level from nuclear rotation~\cite{TownesSchawlow,brown_carrington,kajitaPRA2017} are smaller by approximately $10^3$. They should also largely cancel for transitions that do not change $\Omega$, $J$, and $M_J$ because the nuclear-rotation $g$ factors should be comparable for the two vibrational states.

\section{Prospects}

\subsection{Choice of state and techniques}

The \OOp~molecule is amenable for many two-photon transitions, and there is not an obvious ``best'' choice. While the sensitivity $f_\mu$ increases to a maximum at $v_X = 28$, growing anharmonic effects make its sensitivity only 50\,\% larger than that of $v_X=12$. With a few exceptions like $v_X = 19$, the systematic effects are comparable. The growth of the blackbody shift is not a concern. For example, the $6.4\times10^{-17}$ shift at $v_X=28$ and 300~K means a temperature stability of only 1~K will provide an uncertainty on the BBR shift of $1\times10^{-18}$.

The available laser technology should play a role. The 1063~nm wavelength for $v_X=11$ has multi-watt ytterbium-doped fiber amplifiers available. The transitions to $v_X=12-17$ are all within the range of a titanium-sapphire laser. Many transitions are available as diode lasers, some in tapered amplifiers. For dissociative detection, the UV lasers may be more challenging. For some species of co-trapped atomic ions, existing lasers could fill that role. For example, the 235~nm photoionization laser for Be$^+$ is near the 238~nm wavelength to dissociate $v_X=11$. Similarly, the 313~nm Be$^+$ cooling laser could work for $v_X=15$ or 16 (323~nm or 351~nm). Lasers for Mg$^+$ (280~nm cooling, 285~nm photoionization) would also work, but heavier ions' wavelengths may be too long (for example, Ca$^+$ at 397~nm). Actual calculations of photodissociation cross-sections would help in experimental preparations.

Of the systematic effects, the time-dilation and probe-laser AC Stark shifts pose the largest concerns. To achieve the ultimate accuracy will require a very well compensated trap. Early experiments, however, need not have particularly good trap compensation or even confine the ions near the rf null. For example, the existing molecular-beam accuracy ($6\times10^{-14}$) could be matched with ions in a 3D crystal with approximately \SI{45}{\micro\meter} radius (for our typical trap parameters above). Even achieving $1\times10^{-16}$ only requires compensation that keeps $\mathcal{E}_{\rm DC}\lesssim 1~{\rm V/m}$, which is straightforward in larger traps.

The AC Stark shift of the probe laser will typically be much larger than the target accuracy and thus must be compensated. For example, with $\Omega_R/(2\pi)=10~{\rm Hz}$ for the $v_X = 11$ transition, the probe light shift is $-1\times10^{-12}$ in frequency and $3\times10^{-12}$ in $\Delta\mu/\mu$. For a target accuracy of $6\times10^{-14}$, this shift can be calibrated and the optical power kept stable to better than 2\,\%. Higher accuracy experiments should use a composite pulse sequence immune to the shift~\cite{yudinPRA2010,yudinPRAppl2018,sannerPRL2018,zanonwilletteRPP2018}. In ref.~\cite{huntemannPRL2012}, such a scheme yielded four orders of magnitude suppression of shifts of size $2\times10^{-12}$ and $1\times10^{-13}$ on the Yb$^+$ octupole transition. Reducing the Rabi frequency would also reduce the shift, though it would require longer coherence times for the laser.

Because of the intrinsically narrow lines of these transitions, statistical uncertainty will average quickly to the systematic limits as long as the majority of the time can be spent probing the ions. In general, measuring $N$ ions over time $\tau$, where the linewidth is $\gamma$ (for example in hertz) and the single probe time is $1/\gamma$, should yield a statistical uncertainty of~\cite{hannekePRA2016}
\begin{equation}
	\frac{\delta \mu}{\mu} = \sqrt{\frac{\gamma}{N \tau}}\frac{1}{f_\mu}, %
\end{equation}
where we have assumed the signal-to-noise is limited by quantum projection noise~\cite{itanoPRA1993}. For example, with a 10~Hz linewidth, the $v_X=11$ transition could be measured with statistical precision $\delta\mu/\mu\sim5\times10^{-15}/\sqrt{N(\tau/{\rm s})}$.

In order to spend the majority of the time probing the ions, there is a trade-off in state preparation and detection schemes. Lower-precision experiments that can use multiple ions in a 3D crystal can use the somewhat simpler but destructive dissociation and time-of-flight mass spectrometry detection scheme. Higher-precision experiments in better-compensated traps will likely use quantum-logic detection schemes. Loading by REMPI is an efficient way to prepare the vibrational state of the ion. One-color REMPI uses one fewer laser, but at the expense of some ions in the wrong rotational state. This may be appropriate for multi-ion experiments, but at the cost of signal because not all molecules participate in the experiment. For fewer ions or higher efficiency, a second resonant REMPI laser will be preferred. With a quantum-logic detection scheme in place, it can also be used to prepare the initial state. Proposals exist to prepare different rotational states projectively~\cite{leibfriedNJP2012,dingNJP2012}, though an additional femtosecond laser would be required. 

\subsection{Reference transitions} \label{sec:reference}

In order to detect any change in the transition frequency, and thus change in $\mu$, measurements must be compared with a reference that has different $f_\mu$. An obvious choice would be an optical atomic clock~\cite{ludlowRMP2015}. Because these are based on electronic transitions, they have very small $f_\mu$. They are readily compared to the \OOp~transitions by use of a frequency comb. It is possible that an atomic ion present in the same trap could provide the dual role of both clock and coolant/logic ion. Alternatively, a second transition in \OOp~could be employed~\cite{kajitaJPSJ2017}. By use of a co-trapped atomic ion or two optical transitions in the same molecule, both frequencies would be measured in the same electromagnetic environment, which could simplify evaluations of systematics. The cost of using \OOp~for both frequencies is a smaller differential sensitivity to $\mu$-variation.

For lower-precision experiments in the $10^{-14}$ range, microwave or vapor-cell references would suffice. A frequency comb referred to either GPS or a local cesium clock could be used for any of the transitions. This was the approach taken with SF$_6$ in ref.~\cite{shelkovnikovPRL2008}. Iodine vapor cells provide reference lines at room temperature for transitions to $v_X=19-29$, with extensions to $v_X=9-12$ if the probe laser is doubled before the iodine cell~\cite{gerstenkornI2Atlas1978}. Any sensitivity to $\mu$ variation in the microwave or vapor reference would need to be included in the net sensitivity. With the I$_2$ vapor cell, most of the transition energy is an electronic transition from $X\,^1\Sigma_g^+$ to $B\,^3\Pi_{0u}^+$, so $f_\mu$ is typically much smaller than in \OOp.

\section{Conclusions}

Two-photon vibrational transitions in \OOp~provide multiple routes to improved searches for $\mu$ variation. We have presented experimental procedures capable of searching for $\dot{\mu}/\mu$ with the highest precision achieved in molecules with prospects capable of improving current overall limits. The calculated transition dipole moments have allowed us to estimate the two-photon transition rates and electric-dipole-related systematic effects for each overtone. Additional systematic and statistical estimates show long-term promise at the $10^{-18}$ level or better. Before embarking on a full time-variation experiment, additional spectroscopy is needed to reduce the transition uncertainties far below their current level of several gigahertz.

\vspace{6pt} 

\supplementary{The following are available online: RKR potentials; wavefunctions, energies, and rotational constants for 55 vibrational states within $X$ and 31 states within $A$; overlap integrals, energy differences, and transition dipole moments for all pairs of vibrational states.}

\authorcontributions{All authors contributed to the work presented.}%

\funding{This research was funded by the US National Science Foundation grants PHY-1255170 and PHY-1806223.}

\conflictsofinterest{The authors declare no conflict of interest.}
\reftitle{References}

\end{document}